\documentclass[manuscript]{acmart}
\AtBeginDocument{%
  }

\usepackage{arydshln}
\usepackage{graphicx}

\copyrightyear{2026}
\acmYear{2026}
\acmConference[AHs 2026]{The Augmented Humans International Conference 2026}{March 16--19, 2026}{Okinawa, Japan}
\acmBooktitle{The Augmented Humans International Conference 2026 (AHs 2026), March 16--19, 2026, Okinawa, Japan}
\acmDOI{10.1145/3795011.3795065}
\acmISBN{979-8-4007-2351-3/2026/03}

\usepackage{fancybox} 
\usepackage{xcolor}   

\definecolor{lightgraybox}{gray}{0.95}

\newsavebox{\greyboxbox}
\newenvironment{greybox}{%
  \par\medskip\noindent
  \begin{lrbox}{\greyboxbox}%
  \begin{minipage}{\dimexpr\linewidth-2\fboxrule-4\fboxsep\relax}%
}{%
  \end{minipage}%
  \end{lrbox}%
  \begingroup
    \setlength{\fboxrule}{0.4pt}%
    \setlength{\fboxsep}{0pt}
    \cornersize*{8pt}
    \color{gray!40}%
    \ovalbox{%
      {\setlength{\fboxsep}{6pt}%
       \colorbox{lightgraybox}{\usebox{\greyboxbox}}}%
    }%
  \endgroup
  \par\medskip
}

\begin{document}

\title{15 Years of Augmented Human(s) Research: Where Do We Stand?}

\author{Steeven Villa}
\orcid{0000-0002-4881-1350}
\affiliation{%
  \institution{LMU Munich}
  \city{Munich}
  \country{Germany}
}
\email{villa@posthci.com}

\author{Abdallah El Ali}
\orcid{0000-0002-9954-4088}
\affiliation{%
  \institution{Centrum Wiskunde \& Informatica}
    \city{Amsterdam}
    \country{The Netherlands}}
  \affiliation{%
  \institution{Utrecht University}
   \city{Utrecht}
  \country{The Netherlands}
}
\email{aea@cwi.nl}


\renewcommand{\shortauthors}{Villa and El Ali}

\newcommand{\red}[1]{\textcolor{red}{#1}}
\newcommand{\blue}[1]{\textcolor{blue}{#1}}
\newcommand{\purple}[1]{\textcolor{purple}{#1}}
\newcommand{\orange}[1]{\textcolor{orange}{#1}}
    
\begin{abstract}

The Augmented Human vision broadly seeks to improve or expand baseline human functioning through the restoration or extension of physical, intellectual, and social capabilities. However, given the rapid pace of technology development, we ask: what exactly does Augmented Human research involve, what are its core themes, and how has the Augmented Human(s) conference series evolved over time? To answer this, we conducted a scientometric analysis on the past 15 years of the Augmented Human(s) conference (N=735 paper), focusing on: geographical aspects, submissions and citation timelines, author frequency and popularity, and topic modeling. We find that: (a) Number of papers in the conference exhibit a bimodal distribution, peaking in 2015 and 2025, but showing periods of stagnant growth; (b) key topics over time include Haptics, Wearable Sensing, Vision \& Eye Tracking, Embodied Interaction, and Sports / Motion; (c) some seminal papers on AH are not published in AH(s), but rather at related venues (e.g., CHI); (d) the conference has an active Japanese HCI community despite its historical Eurocentric location dominance. We contribute a closer look at the trajectory of the AH(s) field, and raise considerations of definitional and research scope ambiguities given the core problems/enhancements the field seeks to address.




\end{abstract}


\begin{teaserfigure}
    \centering
    \includegraphics[width=0.8\linewidth, trim={2.6cm 1.3cm 3.2cm 2.8cm},clip]{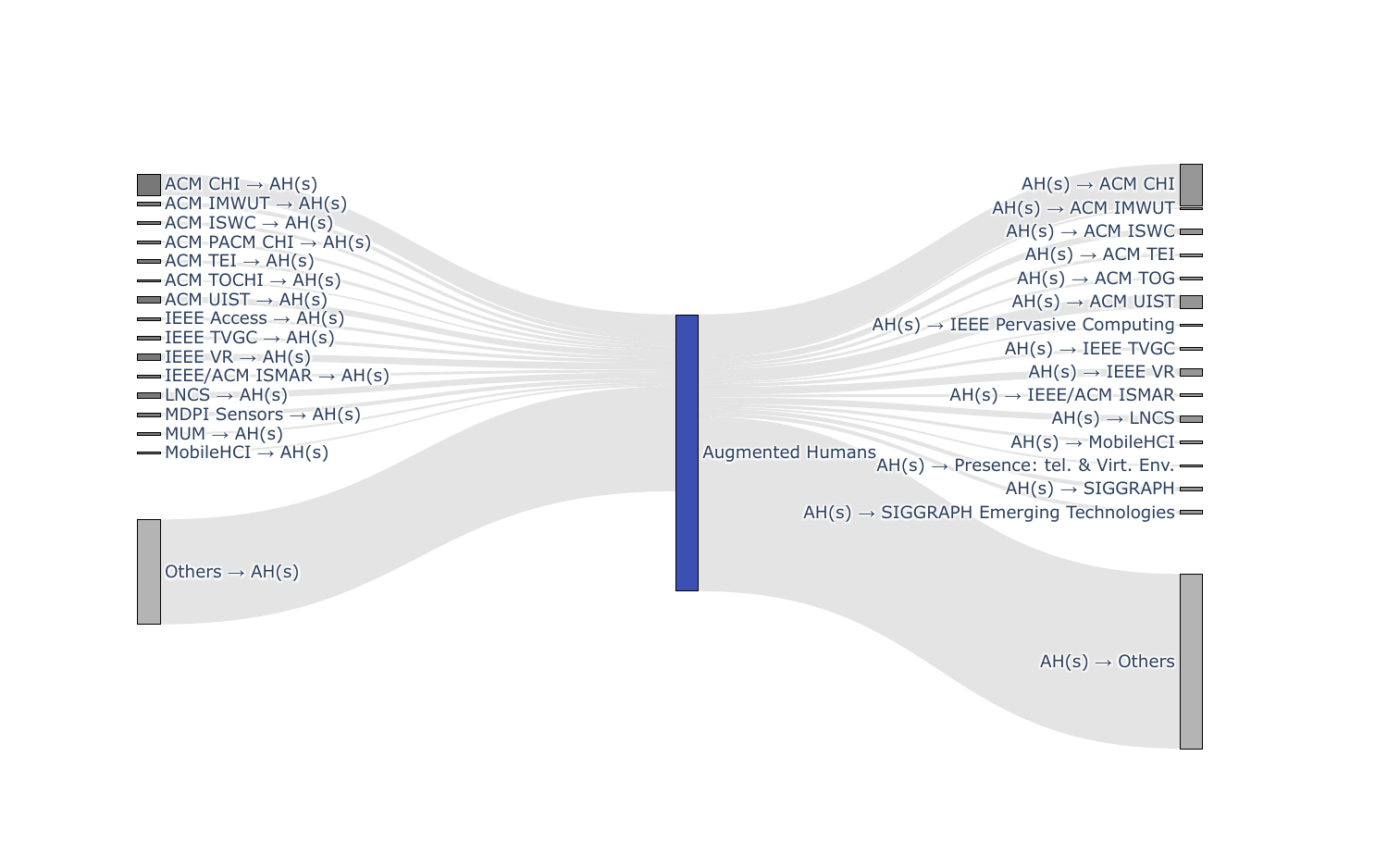}
    \caption{Sankey diagram showcasing the intellectual foundations and external impact of the AH(s) conference. The ACM CHI conference is highly influential in the AH(s) community as foundation, but it is also impacted by AH(s) works.}
     \Description{Sankey diagram showcasing the intellectual foundations and external impact of the AHs conference. The ACM CHI conference is highly influential in the AHs community as foundation, but it is also impacted by AHs works}
    \label{fig:sankey}
\end{teaserfigure}



\maketitle

\section{Introduction}

Augmenting human capabilities with technology has become a key theme in Human-Computer Interaction (HCI). This reflects the ongoing development of systems and technologies that enable integrations between humans and machines \cite{Inami2022jizai,Mueller2020nextsteps-hint,Raisamo2019ah-pastpresent}. Over the past two decades, researchers have explored a wide spectrum of enhancements, ranging from medical prostheses and exoskeletons to perceptual tools such as AR/VR glasses and wearable sensors, all of which aim to restore, supplement, or even exceed human potential \cite{DeBoeck2024HumanAugmentation, Raisamo2019ah-pastpresent}. In the context of HCI, this sub-field is often named as human augmentation (HA), and, in practice, it explores how on- or near-body digital technologies can restore or enhance human sensory, motor, and cognitive capabilities \cite{Villa2023Understanding}. In parallel HCI itself has been evolving in a similar direction, where HCI and related human-factors fields are moving toward an integration of human and artificial intelligence. This has already triggered a debate of whether the future of HCI is human augmentation, or whether human augmentation can stand as its own discipline \cite{Chignell2024evolutionHCI}. Given this, the Augmented Human conference (established in 2010; h5-index: 19, h5-median: 25\footnote{\url{https://scholar.google.com/citations?hl=en&view_op=list_hcore&venue=HqiGvACsawYJ.2025}}) and related venues (e.g., CHI, UIST) have spawned much research, from bionic limbs, brain–computer interfaces, cognitive tools, to embodied extensions of human–computer interaction \cite{Guerrero2022AugmentedHumanity}. In one implementation of this Augmented Human vision, there are efforts to conceptually move from the traditional HCI paradigm toward human–computer integration (HInt) or Augmentation (HA), essentially towards a blend of humans and machines, rather than having machines, computers, or UIs as simple tools in isolation \cite{Mueller2022integratinghumanbodymachine,Mueller2020nextsteps-hint}.


We find that the field of Human Augmentation itself faces conceptual, methodological, and socio-ethical challenges that may complicate its consolidation and responsible advancement \cite{DeBoeck2024HumanAugmentation, Raisamo2019ah-pastpresent, Guerrero2022AugmentedHumanity}. At an anecdotal / observational level, the term “augmented human” has been used inconsistently, overlapping with ``augmented reality,” ``human enhancement,” and even the philosophical view of ``transhumanism,” often without clear boundaries. For example, \citet{Rakkolainen2026Augmentedxtranshuman} finds that although augmented human technologies and transhumanist ideas both aim to extend human capacities, they differ significantly in scope and assumptions (cf., \cite{Barranquero2025Differentiating}). This definitional ambiguity is mirrored in the diversity of topics presented at the Augmented Humans conference\footnote{See e.g., AHs 2026 Call for Papers \url{https://augmented-humans.org/}}, from interface design and wearable systems to ethical, societal and philosophical analyses, all of which can muddle discourse on what constitutes Augmented Humans (AHs) research, and in turn, how progress should best be measured and assessed. Moreover, as augmentation technologies transition from the lab into everyday life \cite{DeBoeck2024HumanAugmentation}, it comes as no surprise that the community would need to address critical issues around privacy \cite{Villa2023Understanding, Guerrero2022AugmentedHumanity}, user autonomy \cite{Cornelio2022ageny-hint}, safety, and the risk of exacerbating a technological divide based on access \cite{Nanayakkara2023assitive}, appearance, or disability \cite{Williams2023cybordassemblages}.


To better understand and realize the promise of augmenting human ability, and to truly delineate what is meant by Augmented Humans, the field may benefit from a more structured articulation of its foundations and empirical contributions. This includes defining what is and is not core AHs (as empirically observed in the conference series), and how this may relate to tangential visions such as transhumanism or cyborg philosophy (e.g., Clark's vision of extended mind \cite{Clark2003cyborgs}). To address this, we conduct a scientometric analysis on the past fifteen years of the Augmented Humans\footnote{Throughout this work, we aim to encompass both communities through the abbreviation of AH(s).} conference (AHs, in plural) and its predecessor Augmented Human conference (AH, in singular). Such analyses provide a quantitative foundation for understanding the evolution and intellectual structure of scientific communities, an activity broadly defined as the science of analyzing science \cite{Henry2007}. Within HCI, several researchers have conducted scientometric and/or bibliometric analyses of conferences (including CHI conference \cite{Bartneck2009}, CSCW \cite{Correia2018}, OzCHI \cite{Mubin2017ozchi}, HAI \cite{Mubin2017hai}), where such analyses typically focus on citations, topic trends, and author frequency and institutions. Our primary research question is: How has the AH(s) field developed over time? More specifically, what are the main topics being presented and published there, what is their impact, and how does the conference series cater to this? To answer this, we employ a mixed approach combining bibliometric analysis, network analysis, and computational text mining, that allows us to analyze publication patterns, scientific influences, key contributors, as well as topic evolution over time. 

\begin{greybox}
\textbf{Clarification:}

\textit{Augmented Human\textbf{s}} (plural) emerged in 2020 following a split from the original \textit{Augmented Human} (singular) community. Both venues co-existed from 2020–2022, after which \textit{Augmented Human} transitioned into the \textit{Augmented Human Research journal}. This paper focuses on the conference and community that evolved into \textit{Augmented Human\textbf{s}}; papers from the 2020–2022 \textit{Augmented Human} conference are analyzed separately for completeness. Accordingly, \textbf{Augmented Human(s)} is used as an umbrella term encompassing both the singular and plural conference trajectories of the field.
\end{greybox}

Our key findings show that: (a) the AH(s) conference publication numbers exhibit a bimodal distribution, with peaks in 2015 and again in 2025, however with some periods of stagnant growth; (b) the dominant research themes were: Haptics, Wearable Sensing, Vision \& Eye Tracking, Embodied Interaction, and Sports / Motion. Whereas Haptics emerged as the most prevalent topic with increasing community interest, Vision \& Eye Tracking showed decline over the years; (c) the highest cited papers on AH(s), some of which are seminal works, are not published in AH(s), but rather at ACM CHI\footnote{\url{https://en.wikipedia.org/wiki/Conference_on_Human_Factors_in_Computing_Systems}} or UIST\footnote{\url{https://en.wikipedia.org/wiki/ACM_Symposium_on_User_Interface_Software_and_Technology}}, indicating considerations of venue impact or prestige; (d) the ACM CHI conference dominates as the most referenced venue, whereas AH(s) has the highest external impact on ACM CHI, UIST, but primarily on a constellation of "other" venues; and (e) the most active contributors to the conference come from the Japanese HCI community, despite the largely Eurocentric location of the conference's history.


Our work serves to help steer the field of AH(s) through not only resolving definitional ambiguities (given entanglements with closely related areas like Transhumanism), but to primarily expand on and raise awareness of the core problems/enhancements the conference seeks to address. In the next sections, we firstly summarize our data collection and data preparation process. Thereafter, we describe the trends found in the data, with a focus on the topical analysis. We then speculate and discuss the potential implications for the AH(s) research community.

\section{Background}
In this section we provide an overview of Augmented Humans research, provide a look into the working definition of Human Augmentation as a field for this paper and highlight the relevance of this scientometric analysis for the Augmented Humans community.

\subsection{Definitions, Visions, and Boundaries of Augmented Humans Research}
\label{sec:def}

The concept of enhancing human abilities through technology has roots tracing back to early visions of man-computer symbiosis \cite{Licklider1960Man} and Douglas Engelbart’s seminal work on augmenting human intellect \cite{Engelbart1962}. This domain, frequently referred to as Human Augmentation (HA) or Augmented Human (AH) \cite{Raisamo2019ah-pastpresent, Guerrero2022AugmentedHumanity, Alicea2018Integrative}, signifies a paradigmatical change within HCI, marked by moving away from the design of interfaces to the integration of technology with the human body \cite{Rekimoto2014NewYou}. This emerging Human-Computer Integration/Human Augmentation paradigm is characterized by a close coupling or fusion of computational and human systems \cite{Mueller2020nextsteps-hint, Alicea2018Integrative, Lee2024Bridging, Nanayakkara2023assitive}, ultimately shifting research focus away from "How do we interact with computers?" toward "How are humans and computers integrated?" \cite{Mueller2020nextsteps-hint}. Based on prior works \cite{Guerrero2022AugmentedHumanity,DeBoeck2024HumanAugmentation,Rakkolainen2026Augmentedxtranshuman,Rekimoto2014NewYou,Rekimoto1995augmentedcomputer} and our own observations, we provide the following working definition of the field:

\begin{greybox}
\textbf{Definition:}

\textit{The field of Augmented Humans seeks to \textbf{improve or expand baseline human functioning through the restoration or extension of physical, cognitive, affective, sensory, and social capabilities}. This involves developing practical technology, spanning sensors and on-body devices, implants, brain-computer interfaces, (medical) prostheses and exoskeletons, and perceptual augmentations such as AR/VR glasses, tactile interfaces, and wearable sensors.}

\end{greybox}


Indeed, augmentations can vary in their degree of integration with the user, ranging from external tools to systems experienced as intuitive or self-like (cf., \cite{Clark2003cyborgs}). Their effectiveness is often studied with respect to \textbf{user agency, embodiment, identity, control distribution, and/or interaction with the environment}. Despite such degrees of tool separation, it is becoming a more widely accepted notion in HCI that HA technologies primarily use near-body digital systems, serving as the user’s subordinate rather than assuming full autonomous control over the task \cite{Villa2023Understanding}.


This pragmatic view of HA can be sharply distinguished from Transhumanism (TH), a closely related philosophical and cultural framework \cite{Rakkolainen2026Augmentedxtranshuman, Popoveniuc2022Transhumanism}. While both movements champion using technology to enhance human abilities \cite{Rakkolainen2026Augmentedxtranshuman}, TH is an ideological and speculative endeavor focused on achieving radical change, ultimately aiming to transcend biological limitations and evolve into a new species, namely Homo Excelsior \cite{Barranquero2025Differentiating, Rakkolainen2026Augmentedxtranshuman}. By contrast, HA operates on a near-future timeline, focusing largely on incremental improvements using currently available, well-regulated engineering and medical tools \cite{Rakkolainen2026Augmentedxtranshuman}. This last aspect, is indeed frequently overlooked in the sense that for a technology to be considered HA, should, in some way, improve human performance, augment the \textbf{human}, not only the environment, not only the interaction. However, despite the efforts to clearly delineate these boundaries, the field of Augmented Humans research remains young, and its core concepts and definitions can be criticized as both broad and vague \cite{Raisamo2019ah-pastpresent, Guerrero2022AugmentedHumanity,Villa2023Understanding}. Such lack of clarity makes it difficult to characterize what truly constitutes core Augmented Humans research, which can potentially lead to misuse of terminology \cite{Guerrero2022AugmentedHumanity}. As such, resolving this definitional ambiguity becomes essential, especially given that the HA focus represents a critical (cf., HCI grand challenges \cite{Stephanidis2025Revisited}), and sometimes historically neglected path for HCI, which would strongly benefit from foundational principles of increasing human capacity and capability, or through assistive augmentation \cite{Nanayakkara2023assitive}, through symbiotic relationships between humans and technology \cite{Chignell2024evolutionHCI}. Given the foregoing, our work aims to further elucidate the scope and works (in the conference series) driving the field of Augmented Humans, through a scientometric lens.

\subsection{Role of Scientometric Analysis for Research Progress}
\label{sec:scientometric_analysis}


By systematically studying publication patterns, citation flows, and metadata, these bibliometric methodologies offer a robust, data-driven perspective essential for moving beyond anecdotal accounts within fast developing fields \cite{Bartneck2009, Henry2007}. Such studies can help provide novice researchers with a reliable roadmap to landmark research and current topical trends, while at the same time offering a global overview necessary to clarify intuitions about the field’s structure and dynamics \cite{Henry2007}. By systematically collecting and analyzing publication data, researchers can gauge the growth and impact of conferences, evaluate the relative quality of papers, and establish essential historical context for future work \cite{Mubin2017hai, Mubin2017ozchi, Nichols2015}. Within HCI, several researchers have conducted scientometric and/or bibliometric analysis of
conferences, where such analyses typically focus on citations, topic trends, and author frequency and institutions. This includes the CHI conference
\cite{Bartneck2009}, CSCW \cite{Correia2018}, OzCHI \cite{Mubin2017ozchi}, HAI \cite{Mubin2017hai}, and so on. Moreover, geographical aspects are important for understanding where communities gather. As such, we believe it is extremely valuable to study and reflect on the publication data of a specialized conference such as Augmented Humans.


We believe this is needed for the Augmented Humans conference, given the field’s intellectual core is undergoing rapid growth and development, primarily due to incredible advances in technology (cf., AI's fast growth alongside HCI \cite{grudin2010ai}). As highlighted earlier, the history of HCI shows that the discipline is intrinsically multidisciplinary, marked by a ``nomadic” tendency driven by rapid technological shifts (cf., \cite{myers1996brief}) that result in knowledge that is highly contextual rather than perhaps universal \cite{Liu2014Mapping, Gurcan2021}. This inherent diversity, while certainly a unique point and strength, can pose challenges for assessing manuscript quality through peer-review, or aligning research efforts across otherwise disparate paradigms \cite{Bartneck2009, Liu2014Mapping, Correia2018}. As such, one key observation across previous HCI research is the absence of strong, cohesive ``motor themes" \cite{Kostakos2015BigHole}, which are well-established areas of knowledge applicable across new technologies \cite{Liu2014Mapping}. To that end, we believe a scientometric review is necessary to help demarcate the central, core, and backbone research themes currently driving Augmented Humans research, and on a practical level, identify and characterize what kind of work gets accepted (and cited) there.







\section{Scientometric Analysis}

This scientometric analysis examines 15 years of the Augmented Human(s) Conference, analyzing publication patterns, scientific influences, key contributors, and topic evolution. We employed a mixed approach combining bibliometric analysis, network analysis, and computational text mining.

\subsection{Data Collection and Preprocessing}

In order to build the core content for the scientometric analysis we followed a series of steps, specified as follows:

\subsubsection{Corpus Assembly}

We collected the complete corpus of the Augmented Human International Conference (2010–2022) and the Augmented Humans International Conference (2020–2025) from the ACM Digital Library. For each conference year, we extracted bibliographic metadata including publication year, keywords, venue location, authors, titles, URLs, and abstracts. 

\subsubsection{Citation Data}

We retrieved citation data using two APIs: Crossref provided aggregate metrics including reference counts and citation counts for each paper. Semantic Scholar supplied detailed information about citing papers, including DOIs and publication years, enabling more granular citation network analysis.

\subsubsection{Text Preprocessing}

We implemented a multi-stage preprocessing pipeline to prepare documents for topic modeling: First we eliminated copyright notices, DOI references, ACM reference format blocks, and venue location information, then we isolated scientific content by extracting text between ABSTRACT and REFERENCES sections when present, or removing all content following the REFERENCES heading, finally we performed a text normalization by converting Unicode ligatures, standardized quotation marks, removed hyphenation artifacts from line breaks, and eliminated lone surrogate codepoints.

\subsection{Analysis Methods}

We conducted a series of analysis to form an overarching view of the Augmented Humans community, these analysis are the following:

\subsubsection{Local Referents}

We constructed a directed citation graph to identify influential papers within the conference community. Nodes represent conference papers, and edges indicated citation relationships between them. For each paper, we calculated the number of internal citations (citations from other papers within the same conference corpus) and ranked papers by this metric to identify the most influential works.

\subsubsection{External Referents}

To understand broader citation patterns, we aggregated all citations across the corpus and ranked the most frequently cited external works. This analysis revealed which papers outside the conference had the greatest influence on the Augmented Humans community.

\subsubsection{External Impact Analysis}

We analyzed the venues where papers citing Augmented Humans research were published. Using the Crossref API, we retrieved venue information for each citing paper, including journal or conference names, publishers, publication types, and event details. We aggregated this data to identify which conferences and journals most frequently cited Augmented Humans work, providing insight into the conference's external reach and impact.

\subsubsection{Reference Analysis}

To understand the intellectual foundations of the conference, we examined reference lists of all papers in our corpus. For each paper, we retrieved its references from Crossref and obtained venue information for works with valid DOIs. This analysis revealed the publication types, venues, and conferences that AH(s) authors most frequently cited, illuminating the interdisciplinary connections and influences shaping the community.

\subsubsection{Topic Modeling}

\paragraph{Model Architecture:} We employed BERTopic \cite{grootendorst2022bertopic}, a transformer-based topic modeling framework, to identify and track research themes. Document embeddings were generated using the all-MiniLM-L6-v2 sentence transformer model, producing 384-dimensional dense vector representations. We accelerated computation using CUDA-enabled GPU processing.

\paragraph{Dimensionality Reduction and Clustering: } We applied UMAP (Uniform Manifold Approximation and Projection) for dimensionality reduction with the following parameters: 5 components, 15 neighbors, minimum distance of 0.0, and cosine similarity metric. Clustering was performed using HDBSCAN (Hierarchical Density-Based Spatial Clustering of Applications with Noise) with a minimum cluster size of 25 documents, Euclidean distance metric, and the Excess of Mass (EOM) cluster selection method with probability estimation enabled.

\paragraph{Topic Representation: } We generated topic representations using a custom CountVectorizer configured with domain-specific stop words to filter common HCI terminology (e.g., "participant", "user", "study", "task", "method", "result") and venue-related terms. The vectorizer extracted n-grams ranging from unigrams to trigrams, requiring terms to appear in at least 2 documents. To improve interpretability, we applied two representation models sequentially: First, Part-of-Speech filtering for noun phrases, adjective-noun combinations, and proper noun sequences. Then, Maximal Marginal Relevance (MMR) \cite{Carbonell1998MMR} for selecting the top 10 most representative and diverse keywords. Following initial extraction, we applied automatic topic reduction using BERTopic's similarity-based merging to consolidate semantically related topics.

\paragraph{Temporal Analysis}

To examine topic evolution, we filtered documents with valid year metadata and excluded outlier assignments. We applied BERTopic's topics-over-time method with both evolution tuning and global tuning enabled to track topic frequency changes across years. Based on top keywords and representative documents, we manually labeled the primary topics. Topic prevalence was calculated as the percentage of corpus documents assigned to each topic per year.
\begin{figure*}[tbh]
    \centering
    \includegraphics[width=1\linewidth]{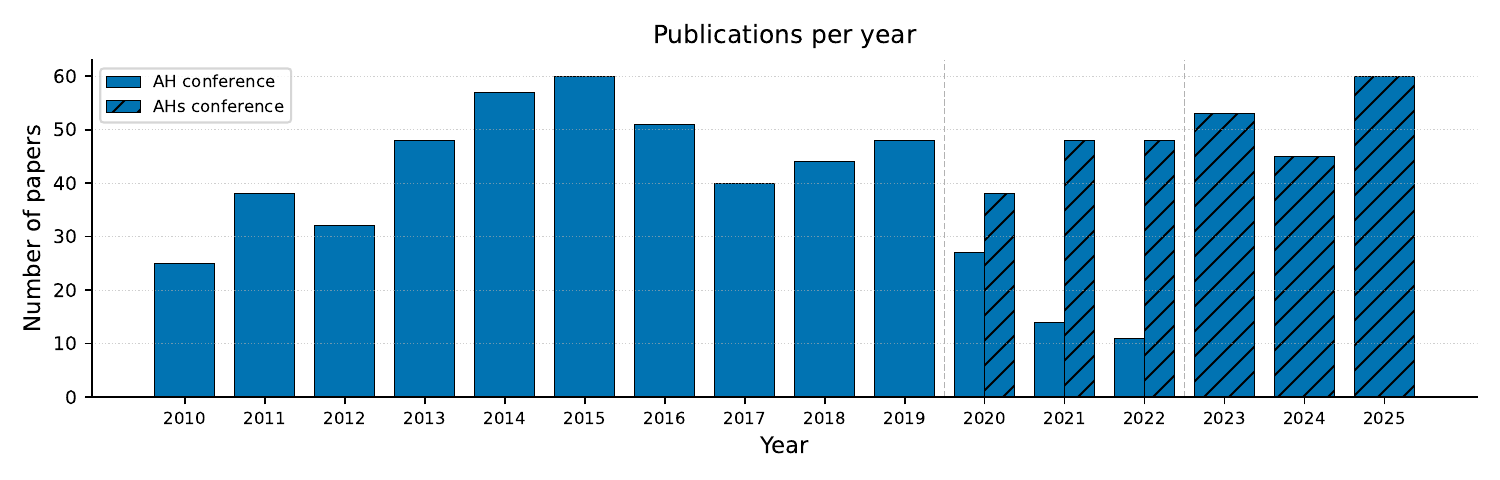}
    \caption{Number of papers published per year. The diagonal lines represent the conference years after the split to Augmented Humans. Dashed lines demarcate the period where both conferences co-existed, until Augmented Human switched to journal format.}
    \label{fig:papersperyear}
\end{figure*}

\section{Results}

We processed a total of 787  papers across 15 years of the conference. In this section, we report the results of the scientometric analysis. We first report the number of publications over time, then the most productive contributors to the conference, followed by the topic modeling analysis. Subsequently, we report the internal and external citation networks, the external impact analysis, and the reference analysis.

\subsection{Conference Metrics}

\begin{table}[t]
  \centering
      \caption{Summary of Augmented Humans conferences.}
  \begin{tabular}{llll}
    \toprule
    Edition & Location    & Acronym  & \footnotesize Accept Rate\\
    \midrule
    2010 & Meg\`{e}ve, France           & AH 10 & 54\% \\
    2011 & Tokyo, Japan                 & AH 11 & - \\
    2012 & Meg\`{e}ve, France           & AH 12 & - \\
    2013 & Stuttgart, Germany           & AH 13 & 71\%\\
    2014 & Kobe, Japan                  & AH 14 & - \\
    2015 & Singapore, Singapore         & AH 15 & - \\
    2016 & Geneva, Switzerland          & AH 16 & 15\%\\
    2017 & Silicon Valley, California, USA  & AH 17 & - \\
    2018 & Seoul, Republic of Korea     & AH 18 & - \\
    2019 & Reims, France                & AH 19 & 49 \%\\\hdashline[0.5pt/2pt]
    2020 & Kaiserslautern, Germany      & AHs'20 & - \\
    2020 & Winnipeg, Canada             & AH 20 & - \\
    2021 & Rovaniemi, Finland           & AHs'21 & 43\%\\
    2021 & Geneva, Switzerland          & AH 21 & - \\
    2022 & Kashiwa, Chiba, Japan        & AHs'22 & - \\
    2022 & Winnipeg, Canada             & AH 22 & - \\ \hdashline[0.5pt/2pt]
    2023 & Glasgow, United Kingdom      & AHs'23 & 38\%\\
    2024 & Melbourne, VIC, Australia    & AHs'24 & 43\%\\
    2025 & Abu Dhabi, UAE               & AHs'25 & - \\
    \bottomrule
  \end{tabular}
  \label{tab:ah_summary}
\end{table}

Overall, the conference has published a total of 787 papers across 15 years, with an average of 45.93 (SD = 9.38) papers per year. 
\autoref{fig:papers_year} shows the number of papers published at the Augmented Humans conference across its 15-year history (2010–2025). The conference began modestly in 2010 with 25 papers and experienced steady growth through its first five years, reaching a peak of 60 papers in 2015. Following this peak, the conference maintained a relatively stable output, averaging 44.3 (SD = 10.3) papers per year from 2016 to 2019.
Although there was an initial decline in publication numbers in the period from 2016 to 2020, the numbers started to increase after the conference re-branding in 2020, for this period (starting from 2020), the average number of papers per edition is 48.66 (SD = 6.77), for the newly created AHs conference.

\autoref{tab:ah_summary} lists all conference edition locations and publicly available acceptance rates. Regarding geographic distribution, the conference has been historically Eurocentric, with 8 of the 15 editions hosted in European countries. Asian countries, particularly Japan, have hosted 4 editions (and the upcoming 2026 edition). The remaining three editions were distributed across America (United States), Oceania (Australia), and the Middle East (United Arab Emirates), each hosting once.

\subsection{Author Productivity}

\begin{figure}[tbh]
    \centering
    \includegraphics[width=0.7\linewidth]{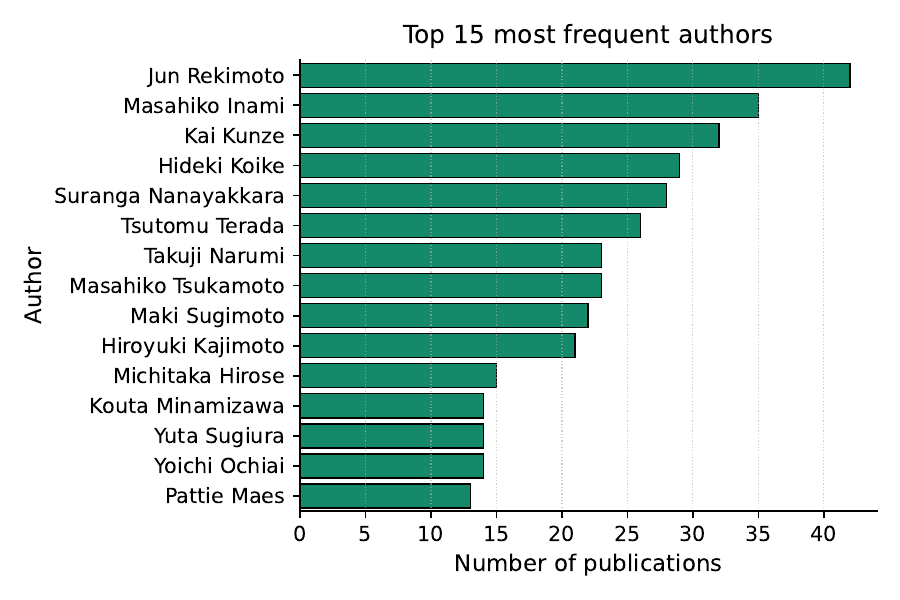}
    \caption{Top 15 Authors historically}
    \label{fig:authors}
\end{figure}

\autoref{fig:authors} presents the top 15 most productive authors in the Augmented Humans conference history. Jun Rekimoto leads with 42 contributions, significantly ahead of other contributors. Masahiko Inami follows with 35 publications, and Kai Kunze with 32 publications, establishing these three researchers as the conference's most prolific contributors.
The distribution reveals a core group of highly engaged researchers from the Japanese HCI community, with 12 of the top 15 authors being affiliated with Japanese institutions. The fourth and fifth most prolific authors, Hideki Koike (29 publications) and Suranga Nanayakkara (28 publications), continue this pattern of sustained contribution. The remaining top contributors show relatively consistent productivity levels, ranging from 11 to 25 publications. An interesting pattern emerging from these trends is the Eurocentric location of the conference versus the Japanese-centric nature of the contributions.

\begin{table*}[bth]
\caption{Summary of identified topics with counts, percentage of corpus, and top words.}
\label{tab:topic_summary}
\begin{tabular}{rcclc}
\toprule
\textbf{Topic Name} & \textbf{Count} & \textbf{Percent} & \textbf{Top Words} & \textbf{Representative Papers} \\
\midrule
\textbf{Haptics} & 116 & 15.8& haptic, tactile, stimulation & \cite{paniaguaTapeTicsFlexibleModular2025, perusquia-hernandezEmbodiedInterfaceLevitation2017a, wittchenDesigningInteractiveShoes2023}\\
\textbf{Wearable Interaction} & 114 & 15.5 & gestures, wearable, sensing & \cite{masaiEyebasedInteractionUsing2020, roddigerPDMSkinOnSkinGestures2020, theissPredictingGraspsWearable2016}\\
\textbf{Vision \& Eye Tracking} & 84 & 11.4 & gaze, vision, eye & \cite{hiroiAdaptiVisorAssistingEye2017, hiroiDehazeGlassesOpticalDehazing2020, kangasFeedbackSmoothPursuit2016}\\
\textbf{Embodied Interaction} & 76 & 10.3 & design, body, virtual & \cite{hatadaDoubleShellfWhat2019, shirotaDesignAlteredCognition2020, tashiroGAuzeMIcrosutureFICATIONGamificationMicrosuture2023} \\
\textbf{Sports / Motion} & 53 & 07.2 & ball, sports, motion & \cite{hamanishiPoseAsQueryFullBodyInterface2020c, itohLaplacianVisionAugmenting2016, sudaPredictionVolleyballTrajectory2019}\\
\bottomrule
\end{tabular}
\end{table*}

\subsection{Dominant Research Topics}
\label{sec:researchthemes}
In our topic analysis, we identified five dominant research themes across the conference corpus: Haptics, Wearable Interaction, Vision \& Eye Tracking, Embodied Interaction, and Sports/Motion.
\textbf{Haptics} emerged as the most prevalent topic throughout the conference's history, encompassing papers on haptic feedback in VR, tactile and vibrotactile stimulation, electrical stimulation, and wearable haptics systems. The second most recurring topic, \textbf{Wearable Interaction}, focused primarily on interaction techniques rather than feedback mechanisms. Papers in this category reported novel input methods using wearables and sensing applications, including activity recognition.
The third topic, \textbf{Vision \& Eye Tracking}, included research on gaze-based interaction, visual displays, eye tracking techniques, and other forms of visual feedback, as well as the use of eyes as an input modality. The fourth topic, \textbf{Embodied Interaction}, comprised papers investigating body representation and embodiment more broadly, including studies of embodiment in VR and the embodiment of robots.
Finally, \textbf{Sports and Motion} encompassed research on various sports (including ball sports and skiing) and physical exertion more generally. \autoref{tab:topic_summary} provides a summary of these topics, including the number of papers assigned to each category and representative example papers.

\begin{figure*}
    \centering
    \includegraphics[width=1\linewidth, trim={0cm 0.4cm 0cm 0cm},clip]{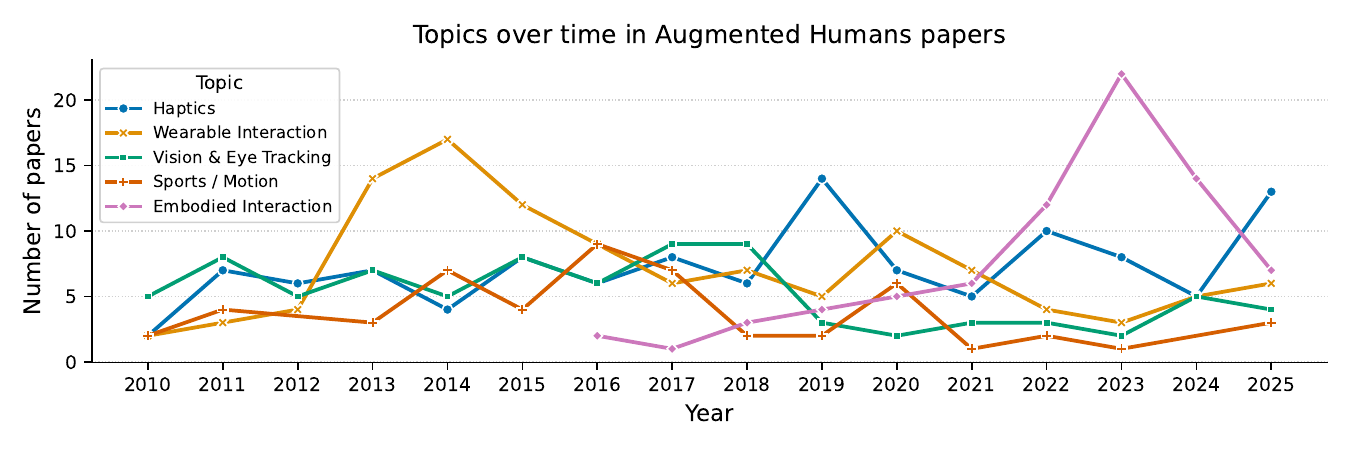}
    \caption{Topic evolution over time: We found five dominant topics in the corpus of the Augmented Humans Conference. The relevance of these topics has shifted over time with Wearable interaction and Vision \& Eye Tracking  being dominant in the early years of the conference, and Embodied Interaction and Haptics being more dominant in recent years.}
    \label{fig:papers_year}
    \Description{Topic modeling.}
\end{figure*}

\autoref{fig:papers_year} illustrates the evolution of the five identified topics across the 15-year history of the conference. The temporal analysis reveals distinct patterns in research focus over time: The conference has evolved from an initial focus on wearable input technologies toward a more diverse research landscape, with haptic feedback and embodied experiences dominating in recent years. \textbf{Wearable Interaction} dominated the early years of the conference (2010-2016), with a notable peak in 2014 at 17 papers, reflecting the strong initial focus on wearable sensing and input techniques during the emergence of wearable computing technologies.
Following this period, \textbf{Haptics} gained prominence and became increasingly central to the conference, particularly from 2017 onward. This topic showed sustained engagement with a significant spike in 2019 (14 papers) and again in 2025 (13 papers), establishing haptics as a core and enduring research area within the AH(s) community.
\textbf{Vision \& Eye Tracking} maintained relatively consistent presence throughout the conference's history, with moderate fluctuations between 5-9 papers annually. However, this topic showed declining attention in recent years, suggesting a potential saturation or shift in research interest.
The most dramatic change occurred with \textbf{Embodied Interaction}, which remained relatively small since 2016 until 2020 but then experienced a significant growth, peaking at 22 papers in 2023. This surge coincides with increased accessibility of VR technologies and growing interest in virtual embodiment research. The topic showed some decline in 2024-2025 but remained a significant area of focus.
\textbf{Sports and Motion} maintained the most stable trajectory, with consistent but modest representation (typically 2-4 papers per year) throughout the conference's history, indicating a niche but sustained research interest. 

\begin{figure}[tbh]
    \centering
    \includegraphics[width=.6\linewidth]{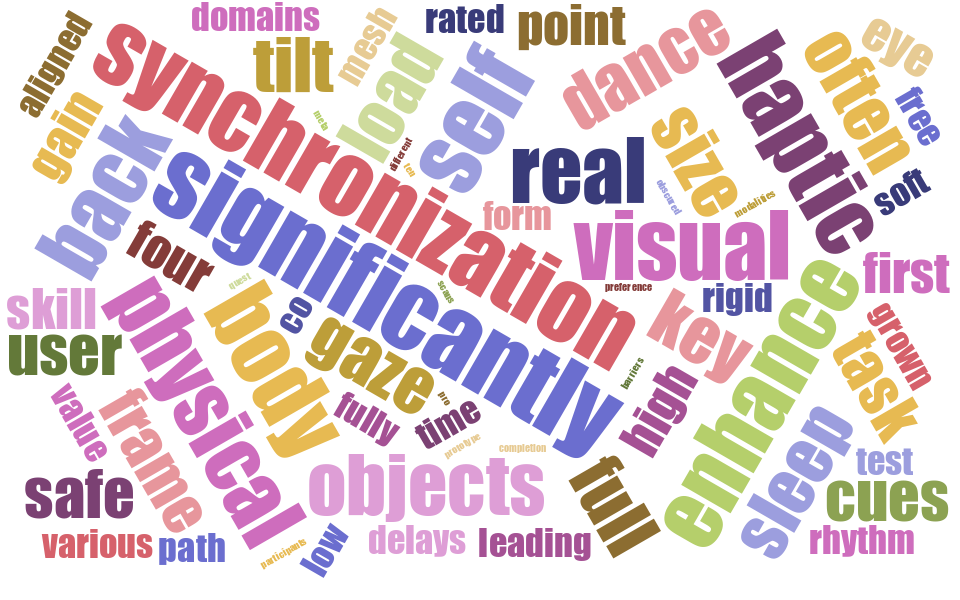}
    \caption{Word cloud over N=733 abstracts.}
    \label{fig:wordcloud}
    \Description{Word cloud}
\end{figure}

To further characterize the type of works published in AH(s), in \autoref{fig:wordcloud} we visualized the works in a word cloud. Pre-processing included removing empty abstract entries (N=2), lowercase conversion, removing elements (punctuation, numerical digits, source tags, common stop words, common academic terms), and filtering entries shorter than three characters. As shown, these dominant themes are well reflected in this visualization.






\begin{table*}[tbh]
\centering
\caption{Top 10 Cited Papers by AH(s)}
\label{tab:top10papers_by_ahs}
\begin{tabular}{@{} l p{10cm} l l l l @{}}
\toprule
\textbf{Venue} & \textbf{Paper Title} & \textbf{Year} & \textbf{$AHs_C$} & \textbf{$ACM_C$} & \textbf{Ref} \\ 
\midrule
CHI & PossessedHand: techniques for controlling human hands using electrical muscles stimuli & 2011 & 20 & 189 & \cite{tamaki2011} \\ 
CHI & Skinput: Appropriating the Skin as an Interactive Canvas & 2010 & 13 & 27 & \cite{harrison2010} \\ 
CHI & The Aligned Rank Transform for Nonparametric Factorial Analyses Using Only ANOVA Procedures & 2011 & 12 & 1853 & \cite{wobbrock2011} \\ 
UIST & OmniTouch: wearable multitouch interaction everywhere & 2011 & 11 & 486 & \cite{harrison2011} \\ 
UIST & MetaArms: Body Remapping Using Feet-Controlled Artificial Arms
 & 2018 & 11 & 95 & \cite{Saraiji2018} \\ 
CHI & Parallel Eyes: Exploring Human Capability and Behaviors with Paralleled First Person View Sharing & 2016 & 10 & 51 & \cite{Kasahara2016paralleleyes} \\ 
CHI & Next Steps for Human-Computer Integration & 2020 & 10 & 155 & \cite{Mueller2020nextsteps-hint} \\ 
AH(s) & Realtime sonification of the center of gravity for skiing & 2012 & 10 & 55 & \cite{Hasegawa2012} \\ 
CHI & Cruise Control for Pedestrians: Controlling Walking Direction using Electrical Muscle Stimulation & 2015 & 10 & 140 & \cite{Pfeiffer2015} \\ 
CHI & iSkin: Flexible, Stretchable and Visually Customizable On-Body Touch Sensors for Mobile Computing & 2015 & 10 & 310 & \cite{Weigel2015} \\ 
\bottomrule
\end{tabular}
\end{table*}

\subsection{Local and External Referents}

We analyzed citation patterns both within the conference community and across the broader research landscape. \autoref{tab:top10papers_by_ahs}, \autoref{tab:top10papers_from_ahs}, and \autoref{tab:top10papers_internal_ahs} present the most influential papers as measured by different citation metrics.

From \autoref{tab:top10papers_by_ahs}, we find that only one paper from the top 10 most cited papers in AH(s) is from AH(s), namely the work by \citet{Hasegawa2012}, where the rest span CHI and UIST. From \autoref{tab:top10papers_from_ahs}, we see that the top 10 most cited papers overall from AH(s), four are about human activity recognition (a common UbiComp topic), three on multimodal input (common to ISWC / UIST), one multimodal feedback, one on XR, and only one that appears to be typical of AH(s), namely work by \citet{Kasahara2014}. From \autoref{tab:top10papers_internal_ahs}, we find that the top 10 internally cited at AH(s), these reflect some of the dominant research topics, where four are on Sports / Motion, four on Vision \& Eye tracking, and two on Embodied Interaction.

\begin{table*}[tbh]
\centering
\caption{Top 10 Cited Papers from AH(s)}
\label{tab:top10papers_from_ahs}
\begin{tabular}{p{10cm} l l l @{}}
\toprule
\textbf{Paper Title} & \textbf{Year} & \textbf{Cites} & \textbf{Citation} \\ 
\midrule
In the blink of an eye: combining head motion and eye blink frequency for activity recognition with Google Glass & 2014 & 135 & \cite{ishimaruBlinkEyeCombining2014} \\
Qualitative activity recognition of weight lifting exercises & 2013 & 112 & \cite{vellosoQualitativeActivityRecognition2013}\\
JackIn: integrating first-person view with out-of-body vision generation for human-human augmentation & 2014 & 111 & \cite{Kasahara2014}\\
HASC Challenge: gathering large scale human activity corpus for the real-world activity understandings & 2011 & 103 & \cite{kawaguchiHASCChallengeGathering2011}\\
EyeRing: a finger-worn input device for seamless interactions with our surroundings & 2013 & 75 & \cite{nanayakkaraEyeRingFingerwornInput2013}\\
Let me grab this: a comparison of EMS and vibration for haptic feedback in free-hand interaction & 2014 & 75 & \cite{pfeifferLetMeGrab2014c}\\
Distributed Metaverse: Creating Decentralized Blockchain-based Model for Peer-to-peer Sharing of Virtual Spaces for Mixed Reality Applications & 2018 & 74 & \cite{ryskeldievDistributedMetaverseCreating2018}\\
TongueBoard: An Oral Interface for Subtle Input & 2019 & 72 & \cite{liTongueBoardOralInterface2019}\\
Wearability Factors for Skin Interfaces & 2016 & 68 & \cite{liuWearabilityFactorsSkin2016}\\
Airwriting recognition using wearable motion sensors & 2010 & 67 & \cite{ammaAirwritingRecognitionUsing2010}\\
\bottomrule
\end{tabular}
\end{table*}

\paragraph{External Papers Most Cited by AH(s) Authors:} The conference's intellectual foundations are heavily rooted in body-based interaction techniques from CHI, particularly electrical muscle stimulation and on-skin interfaces. \textit{PossessedHand} by \citet{tamaki2011} (CHI 2011, 20 citations) and \textit{Skinput} by \citet{harrison2010} (CHI 2010, 13 citations) exemplify this focus. The list is dominated by CHI papers (7 of 10), with methodological work (\textit{Aligned Rank Transform} by \citet{wobbrock2011}) also appearing, suggesting both technical and analytical influences from the broader HCI community.

\paragraph{AH(s) Papers Most Cited Externally:} Augmented Humans has produced substantial external impact, with top papers receiving 67-135 citations. Activity recognition using wearables emerged as the most influential contribution, with \textit{In the blink of an eye} \citet{ishimaruBlinkEyeCombining2014} (2014, 135 citations) leading significantly. The 2013-2014 period was particularly productive, generating 6 of the top 10 most-cited papers. These works have resonated primarily with wearable computing and activity recognition communities.

\paragraph{AH(s) Papers Most Cited Within AH(s):} Internal citations (6-10 per paper) are notably lower than external citations, reflecting the smaller community size. \textit{Realtime sonification of the center of gravity for skiing} by \citet{Hasegawa2012} (2012, 10 citations) serves as both the most internally and externally cited work. Several papers appear across both internal and external citation lists.

\begin{table*}[tbh]
\centering
\caption{Top 10 Internally Cited Papers from AH(s)}
\label{tab:top10papers_internal_ahs}
\begin{tabular}{p{10cm} l l l @{}}
\toprule
\textbf{Paper Title} & \textbf{Year} & \textbf{Cites} & \textbf{Citation} \\ 
\midrule
Realtime sonification of the center of gravity for skiing & 2012 & 10 & \cite{Hasegawa2012}\\
Swimoid: a swim support system using an underwater buddy robot & 2013 & 8 & \cite{ukaiSwimoidSwimSupport2013}\\
Bouncing Star project: design and development of augmented sports application using a ball including electronic and wireless modules & 2010 & 7 & \cite{izutaBouncingStarProject2010}\\
JackIn: integrating first-person view with out-of-body vision generation for human-human augmentation & 2014 & 7 & \cite{Kasahara2014}\\
Laplacian Vision: Augmenting Motion Prediction via Optical See-Through Head-Mounted Displays & 2016 & 6 & \cite{itohLaplacianVisionAugmenting2016}\\
TongueBoard: An Oral Interface for Subtle Input & 2019 & 6 & \cite{liTongueBoardOralInterface2019}\\
Aided eyes: eye activity sensing for daily life & 2010 & 6 & \cite{Ishiguro2010aidedeyes}\\
MultiSoma: Distributed Embodiment with Synchronized Behavior and Perception & 2021 & 6 & \cite{miuraMultiSomaDistributedEmbodiment2021}\\
SpiderVision: extending the human field of view for augmented awareness & 2014 & 6 & \cite{fanSpiderVisionExtendingHuman2014}\\
HoverBall: augmented sports with a flying ball & 2014 & 6& \cite{nittaHoverBallAugmentedSports2014} \\
\bottomrule
\end{tabular}
\end{table*}

\subsection{External Impact and Reference Analysis}
\label{sec:ext_impact}

\autoref{tab:references} presents the citation relationships between AH(s) and external venues, showing both where AH(s) papers are cited (external impact) and which venues AH(s) authors most frequently reference (intellectual foundations). 
\autoref{fig:sankey} visualizes these bidirectional relationships as flows between venues and the conference, with the width of each flow proportional to the number of citations or references. The analysis reveals strong bidirectional connections with the broader HCI and extended reality research communities.

\subsubsection{External Impact:} ACM CHI emerges as the most significant venue citing AH(s) research, with 812 citations representing 12.02\% of all external citations. This substantial citation rate from the premier HCI conference demonstrates the relevance and influence of AH(s) research within the broader human-computer interaction community. As visible in the Sankey diagram, the flow from AH(s) to CHI is the most prominent outgoing connection, reflecting this strong relationship. ACM UIST follows with 228 citations (3.38\%), and IEEE VR with 265 citations (3.92\%), indicating strong connections with specialized communities in user interface software and virtual reality. The conference also shows notable impact in computer graphics venues (SIGGRAPH, SIGGRAPH ET) and ubiquitous computing (ACM ISWC), though these venues cite AH(s) research less frequently. Venues such as ACM IMWUT, IEEE Access, and MDPI Sensors cite AH(s) work but do not appear prominently in AH(s) reference lists, suggesting unidirectional knowledge transfer to these communities, as shown by the one-way flows in the Sankey diagram. The "Others" category accounts for 65.61\% of citations, representing the largest aggregate flow in the visualization and indicating that AH(s) research reaches a diverse array of venues beyond the core HCI and VR conferences.

\subsubsection{Reference Analysis:} ACM CHI also dominates as the most referenced venue, with 1,582 references constituting 14.97\% of all citations made by AH(s) papers. The Sankey diagram shows this as the thickest incoming flow to AH(s), confirming CHI as the primary intellectual hub for the AH(s) community. ACM UIST (497 references, 4.70\%) and IEEE VR (290 references, 2.74\%) follow as key reference sources, maintaining their positions as important foundations for AH(s) research. Interestingly, several venues appear more frequently in AH(s) references than in citations to AH(s) work: SIGGRAPH (158 references vs. 0 citations) and SIGGRAPH Emerging Technologies (145 references vs. 0 citations) are heavily referenced but do not cite AH(s) papers back. These asymmetric relationships are clearly visible in the Sankey diagram as incoming flows with no corresponding outgoing flows, suggesting that AH(s) researchers draw upon computer graphics methods without necessarily contributing back to those venues. Similarly, IEEE/ACM ISMAR (107 references, 1.01\%) is referenced more than it cites AH(s) work. The "Others" category represents 66.98\% of references, slightly higher than the citation percentage and visible as the largest incoming flow in the diagram, indicating that AH(s) authors draw upon an even broader and more diverse set of intellectual sources than the venues that cite their work.

\begin{table}[]
    \centering
    \caption{External Impact and Intellectual Foundations of the Augmented Human(s) Conference: Almost 15\% of the cited works at AH(s) come from the ACM CHI conference and 12.02\% of the citations come from the same conference, showing that AH(s) is strongly tied to this conference}
    \label{tab:references}
    \begin{tabular}{llllrr}
\toprule
Venue & Cites & Referenced & Cites (\%)  & Referenced (\%)  \\
\midrule
\textbf{ACM CHI} & 812 & 1582 & 12.02 & 14.97 \\
\textbf{ACM UIST} & 228 & 497 & 3.38 & 4.70 \\
\textbf{IEEE VR} & 265 & 290 & 3.92 & 2.74 \\
\textbf{LNCS} & 235 & 241 & 3.48 & 2.28 \\
\textbf{ACM ISWC} & 126 & 235 & 1.87 & 2.22 \\
\textbf{SIGGRAPH} & – & 158 & 0.00 & 1.50 \\
\textbf{SIGGRAPH Em. Tech.} & – & 145 & 0.00 & 1.37 \\
\textbf{ACM TEI} & 116 & 118 & 1.72 & 1.12 \\
\textbf{IEEE TVGC} & 126 & 116 & 1.87 & 1.10 \\
\textbf{MobileHCI} & 66 & 115 & 0.98 & 1.09 \\
\textbf{IEEE/ACM ISMAR} & 104 & 107 & 1.54 & 1.01 \\
\textbf{ACM TOG} & – & 91 & 0.00 & 0.86 \\
\textbf{ACM IMWUT} & 153 & 76 & 2.26 & 0.72 \\
\textbf{Presence: Tel. \& Virt. Env.} & – & 71 & 0.00 & 0.67 \\
\textbf{IEEE Pervasive Comp.} & – & 58 & 0.00 & 0.55 \\
\textbf{ACM PACM CHI} & 110 & – & 1.63 & 0.00 \\
\textbf{MUM} & 85 & – & 1.26 & 0.00 \\
\textbf{IEEE Access} & 115 & – & 1.70 & 0.00 \\
\textbf{ACM TOCHI} & 60 & – & 0.89 & 0.00 \\
\textbf{MDPI Sensors} & 147 & – & 2.18 & 0.00 \\\hdashline[0.5pt/2pt]
\textbf{Others} & 4007 & 6667 & 59.32 & 63.09 \\\midrule
\textbf{Total} & 6755 & 10567 & 100 & 100 \\
\bottomrule
\end{tabular}
\end{table}

\subsection{Core Topics of the Augmented Human Conference for Year Range 2020-2022}

In \autoref{tab:topic_summary_AH} we present the summary of the identified stable topics of the Augmented Human (singular) conference for the years ranging from 2020 to 2022, which is the period where both conferences co-existed. It is important to highlight that we did not analyze in detail the fluctuation of these topics, nor the topical exchange across these years between Augmented Human and Augmented Humans. Therefore, the information presented in this section is indicative rather than explanatory of the conference dynamics during this period.

The most dominant and stable topic during these years was \textbf{Conceptual Frameworks} with 13 publications. This cluster includes contributions addressing the design of augmentation systems, theoretical models of human augmentation, communication frameworks, and broader philosophical reflections such as transhumanist perspectives and conceptualizations of technology-mediated human extension.

The second most represented topic was \textbf{Physical Interaction} with 11 papers, including work on actuators, haptic feedback and physically embodied interaction techniques. Papers in this cluster focus on the technical realization and empirical evaluation of bodily augmentation systems, connecting with the previous conference’s years of sustained engagement with hardware-driven and sensorimotor approaches to augmentation.

\textbf{Smartphone Usage} emerged as a distinct topic centered on mobile devices, app-based systems, and behavioral support applications. Contributions in this cluster examine how everyday mobile technologies function as augmentation platforms, particularly in contexts such as health monitoring, reminders, and activity recognition.

Finally, \textbf{Input Techniques} captured research on gesture interaction, touchscreen input, pen-based systems, gaze input, and tactile feedback mechanisms. These works address interaction design at the level of modality and technique, exploring how users control, manipulate, and communicate with augmented systems.

\begin{table*}[bth]
\caption{Summary of identified topics with counts, percentage of corpus, and top words.}
\label{tab:topic_summary_AH}
\begin{tabular}{rcclc}
\toprule
\textbf{Topic Name} & \textbf{Count} & \textbf{Percent} & \textbf{Top Words} & \textbf{Representative Papers} \\
\midrule
\textbf{Conceptual Frameworks} & 13 & 25.0 & design, communication, augmentation & \cite{schmidt_extensions_2020, popoveniuc_transhumanism_2022, angelini_towards_2022} \\
\textbf{Physical Interaction }& 11 & 21.2 & activity, actuators, haptic & \cite{alimardani_deep_2021, schmeier_manipulating_2021, okuno_classification_2022} \\
\textbf{Smartphone Usage} & 8 & 15.4 & smartphone, mobile, apps & \cite{zhang_smart_2020, hasan_exploration_2020, maddahi_caring4dementia_2020}\\
\textbf{Input Techniques} & 7 & 13.5 & gestures, touchscreen, touch & \cite{faleel_user_2020, sun_penshaft_2021, bardot_eyes-free_2020}\\
\bottomrule
\end{tabular}
\end{table*}

\section{Discussion}

In this section we synthesize the main observations resulting from our scientometric analysis, analyze the core themes of the conference, and propose future directions of
the Augmented Human(s) conference.

\subsection{The Augmented Human(s) Conference: Key Observations}


Given our analysis, we distill the following empirical observations: First, we find that AH(s) publication numbers exhibit a bimodal distribution, peaking in 2015 and again in 2025, with some periods of stagnant growth. This may reflect the distinction between the Augmented Humans conference (AHs, in plural) and its predecessor Augmented Human conference (AH, in singular), and their respective research trajectories. Second, we observe that dominant research themes (Sec \ref{sec:researchthemes}) comprised of the following topics: Haptics, Wearable Sensing, Vision \& Eye Tracking, Embodied Interaction, and Sports / Motion. Whereas Haptics emerged as the most prevalent topic with increasing community interest, Vision \& Eye Tracking shows decline. Third, we see that the highest cited papers on AH(s), some of which are seminal or landmark works, are not published in AH(s), but rather at CHI or UIST (e.g., \cite{tamaki2011} or \cite{Saraiji2018}). This may be due to considerations of venue prestige (cf., \cite{Maggio2024_BestHomeForThisPaper}). 

Fourth, regarding citation impact, we find that only one paper from the top 10 most cited papers in AH(s) is from AH(s), namely the work by \citet{Hasegawa2012}, where the rest span CHI and UIST. From the top 10 most cited papers overall from AH(s), four are about human activity recognition (a common UbiComp topic), three on multimodal input (common to ISWC / UIST), one multimodal feedback, one on XR, and the work of \citet{Kasahara2014}, which most typically represents AH(s) research. From the top 10 internally cited at AH(s), these reflect some of the dominant research topics, where four are on Sports / Motion, four on Vision \& Eye tracking, and two on Embodied Interaction. Fifth, we find that ACM CHI dominates as the most referenced venue, whereas AH(s) has the highest impact (with respect to citations) partially on ACM CHI and UIST, but most largely on a constellation of `other' venues. Sixth, we observe that the most active contributors to the conference come from the Japan-based HCI community, which highlights an interesting pattern where the largely Eurocentric location of the conference witnesses a high amount of contributions from Japanese institute affiliations. Lastly, an interesting pattern emerging from these trends is the Eurocentric location of the conference versus the Japanese-centric nature of the contributions.



\subsection{Does the Augmented Human(s) Field have a Core Theme?}

By now it has become clear that HCI is largely an applied and inherently interdisciplinary field, and largely defined by the relationships it creates between disciplines rather than punctuating a singular intellectual foundation \cite{Oulasvirta2016HCI, Blackwell2015Filling}. Are we seeing the same pattern for Augmented Humans research, and if so, does this affect the field's core theme(s)? To answer this, it is perhaps worth first revisiting the status of HCI, and how the field consistently grapples with challenges to its intellectual status and efficiency \cite{Beck2017hcibigquestions}. Throughout the last decade, several critical perspectives have emerged, and identified a persistent so-called "big hole" in HCI research \cite{Kostakos2015BigHole, Blackwell2015Filling}, one that is largely characterized by a lack of a solid intellectual or methodological core. This observation runs parallel to our own and others' observations that the field tends to sometimes chase technological trends instead of pursuing deep, long-term themes, or ``big questions" \cite{Beck2017hcibigquestions, Oulasvirta2016HCI}. This lends itself to disciplinary fragmentation \cite{Oppenlaender2025Keeping, Liu2014Mapping}, which ends up boosting work that is more incremental and context-dependent over cumulative theoretical progress \cite{Wobbrock2016Research, Oulasvirta2016HCI}. Recent bibliometric analyses suggests that the disruptiveness of HCI research, specifically its ability to introduce paradigm-shifting knowledge, is sharply decreasing \cite{Chen2025decreasingdisruptivenesshci} -- and falling at a faster rate than the general decline seemingly seen across science. This is potentially due to increasing consensus on conventional norms that penalize unconventional studies \cite{Chen2025decreasingdisruptivenesshci, Oppenlaender2025Keeping}. However, while HCI’s disruptiveness may appear to be declining, based on our analysis we still find that AH(s) (even if viewed as a sub-field of HCI) encourages speculative, high-risk, and functionally unusual prototypes that can challenge our existing notions of human capability.

Reflecting further, we believe that this persistent disciplinary fragmentation and struggle to consolidate knowledge highlights the necessity of not only establishing strong, focused research communities united by a core theme, but also reflected in a living (and ever-evolving) HCI educational curricula (cf., \cite{churchill2015future,Hornbaek2025rethinkingeducation}). Consider that by organizing around the goal of augmenting human capabilities and designing technology for harmonious integration \cite{Chignell2024evolutionHCI}, whether on- or near-body, we find that the AH(s) community can similarly run the risk of losing sight of crucial HCI grand challenges (cf., human-technology symbiosis \cite{Stephanidis2025Revisited}. However, unlike the fragmentation seen in HCI, we believe AH(s)' emphasis on enhancing / restoring capabilities helps provide the field with perhaps a clearer organizing principle. Based on our analysis, we find that AH(s) does have a core theme, and this is evident when humans merge with machines for the purpose of enhancing human functioning. For example, consider HA factors relating to agency and autonomy \cite{Bennett2023agency,Cornelio2022ageny-hint} (e.g., boosting reaction time performance \cite{Kasahara2019}), virtual co-embodiment \cite{Fribourg2021coembodiment,Venkatraj2024coembodiment}, or shared augmented vision \cite{Kasahara2014}), but also other wild new ideas published in AH(s) which may as of yet not have seen wider recognition (cf., Sec \ref{sec:ext_impact}). Examples include Living Bits \cite{Pataranutaporn2020livingbits}, the JIZAI body \cite{Inami2022jizai}, Aided Eyes for memory enhancement \cite{Ishiguro2010aidedeyes}, flying sports \cite{Higuchi2011flyingsports}, bionic vision using machine learning \cite{Han2021bionicvision}, wearable reasoning assistance \cite{Danry2020wearabelreasoner}, controlling multiple robotic arms \cite{Kazuma2022parallelpingpong}, wearable haptics for collaboration \cite{Matsuda2020hapticpointer}, or perception modification through AR-based reality alterations \cite{Yokoro2024decluttAR,Knierim2020speedofreality}. Across these examples, we find that a coherent theme emerges: AH(s) research seems to systematically explore new forms of human–machine integration in the sensorimotor loop, aiming to expand human action capabilities through sensory and perceptual amplification / modification / diminishment. Moreover, we believe the aforementioned works constitute a unique aspect of AH(s) research, of putting forward brave new ideas that remain grounded in HA, rather than lofty transhumanism goals or common HCI interaction techniques. While there may be overlap with technical HCI and/or AR/VR communities, these types of works help establish a core theme, despite our analysis showing a more scattered topic distribution and unsteady citation counts across the years. 









\subsection{Where Do We Go From Here?}
\label{sec:future}

Our work raises several considerations regarding the future direction of the Augmented Human(s) conference:

\textbf{Intermitted checkpoints to clarify conference vision and scope.}  
The recent calls for papers emphasize physical, cognitive, and perceptual augmentation of humans through digital technologies, while also highlighting societal-scale impact and drawing on the historical lineage of augmenting human intellect. Despite this broad framing however, the relationship between the stated scope of AH(s) and the work actually published remains under-defined. While the range of topics listed as relevant to the conference span multiple established HCI subdomains, it offers limited guidance on what distinguishes an AH(s) contribution from related research at venues (e.g., CHI, UIST, ISWC). As such, intermittently revisiting how the call for papers evolves over time, and what it foregrounds as central concerns versus peripheral interests, may help create checkpoints that can help clarify whether the community is converging on a shared vision of augmentation or whether its breadth obscures the field’s distinctive lens and aims.

\textbf{Revisiting definitions of human augmentation.}  
A closely related question concerns how augmentation itself is defined and/or conceived within the AH(s) community. Early conceptions of human augmentation (e.g., Engelbart’s notion of amplifying human intellect \cite{Engelbart1962}) focused primarily on cognitive extension through interactive systems. However, we find that much of the work presented at AH(s) emphasizes embodied, perceptual, and sensorimotor integration between humans and machines. Whether these perspectives represent complementary interpretations of a common concept (cf., \cite{Mueller2020nextsteps-hint}), or reflect different assumptions about what augmentation entails, remains open for discussion. As such, continually re-engaging with historical definitions of augmentation may help further specify the conceptual boundaries of the field.

\textbf{Knowledge accumulation and methodological expectations.}  
Beyond conceptual framing, our analysis raises practical questions about how knowledge in AH(s) accumulates over time. Many contributions rely on bespoke hardware, narrowly scoped use cases, and/or short-term user evaluations, which can make comparison, replication, and extension challenging. This further complicates the formation of a cumulative research trajectory, particularly for systems that aim to alter perception, action, or skill acquisition beyond brief experimental contexts. Here we posit that rather than positioning AH(s) as independent from HCI, it may be helpful to articulate methodological expectations (e.g., open science approaches) that acknowledge the exploratory and high-risk nature of augmentation research while still supporting longer-term insight through longitudinal, in-the-wild studies.

\textbf{Human agency in AI-mediated augmentation.}  
Recent advances in machine learning and generative artificial intelligence further complicate the development of augmentation systems. As such systems increasingly rely on predictive models and adaptive behavior, the relationship between human agency and machine autonomy becomes less clear. For AH(s), this raises questions about how the human sensorimotor loop is reshaped when intelligence is distributed across human and computational components (e.g., parallel running agents), and how notions of control, responsibility, and dependence should be interpreted in this context.

\textbf{Assistive technologies, inclusion, and social impact.}  
Finally, the future of AH(s) research cannot be separated from its broader social implications. While much work in the community focuses on enhancement beyond normative baselines, the relationship between AH(s) and assistive (cf., \cite{Nanayakkara2023assitive,Tan2025assistiveaug} or accessibility-oriented (cf., \cite{Williams2023cybordassemblages,zhou2025augmented}) research can further flourish, particularly given the presence of established communities such ACM SIGACCESS Conference on Computers and Accessibility (ASSETS)\footnote{\url{https://en.wikipedia.org/wiki/SIGACCESS}}. At the same time, augmentation technologies raise concerns about unequal access and the potential amplification of social disparities. Addressing these issues may require AH(s) researchers to engage more explicitly with questions of inclusion, accessibility, and the long-term societal consequences of deploying augmentation technologies beyond research settings -- such social issues may go well beyond assessment of technical risks and alignment.

\subsection{Limitations and Future Work}

We note that our study provides analysis and interpretation that is inherently limited by the scope of 15 years of the AH(s) conference series. This excludes not only venues that may have HA work (e.g., CHI, UIST), but also the journal Augmented Human Research\footnote{\url{https://link.springer.com/journal/41133}}. While this was a deliberate choice, we acknowledge that this may pose an incomplete view of the HA field. Similarly, we do not consider the myriad of works that pertain to Transhumanism, whether from a medical or philosophical standpoint. Furthermore, our analysis is by definition limited to what was published at AH(s), which intrinsically exhibits a survivorship bias -- there could have been many works that are seemingly far more related to the AH(s) vision and conference topics, but did not make it through. Nevertheless, we do believe that our analysis provides a representative characterization of the field, where it stands, and insights that may help steer its future.


\section{Conclusion}

With the rapid pace of technology development, our work sought to answer what exactly the field of Augmented Human(s) involves, focusing on its core themes and how these evolved over time. To answer this, we conducted a scientometric analysis on the past 15 years of the Augmented Human(s) conference series, focusing on: geographical aspects, submissions and citation timelines, author frequency and popularity, and topic modeling. Our key findings showed that the number of papers in the conference exhibit a bimodal distribution, with peaks in 2015 and 2025, however also showing periods of stagnant growth. We found that key topics over time include Haptics, Wearable Sensing, Vision \& Eye Tracking, Embodied Interaction, and Sports / Motion, and these largely cover some of the internally most cited works. We also find that some seminal papers on HA are not published in AH(s), but rather at related venues (e.g., CHI, UIST). Lastly, we observe that the conference has an active Japanese HCI community while its geographic history shows Eurocentric dominance. Our work contributes a closer look at the trajectory of the AH(s) field, where we hope it raises meaningful considerations regarding definitional and research scope ambiguities given the core problems/enhancements of human capability that the field focuses on.

\section*{Positionality and AI Usage Disclosure Statement}

We recognize that author positionality shapes the perspectives and interpretations in this paper~\cite{olteanu2023responsibleairesearchneeds}. We are two researchers situated in Germany and the Netherlands, with backgrounds in Cognitive Science, HCI, and Engineering. Our diverse perspectives informed our research framing and analysis. In the spirit of augmenting human capability, we used LLMs (GPT-5, Claude Sonnet 4, Gemini 3 Pro) to gather unknown references for related work, provide feedback on section structures and argumentation, and support with data analysis and plotting functions. All analytical decisions and interpretive claims, specifically all interpretation, reflection, and discussion, were human-only. We disclose our AI usage (cf.,~\cite{ElAli2024aidisclosure}) in efforts to ensure transparency in AI-mediated scholarship within HCI~\cite{Elagroudy2025lwms} and beyond.

\begin{acks}
\noindent
\begin{minipage}[t]{3.2cm}
  \centering
  \fbox{\raisebox{-2.0\baselineskip}{\includegraphics[width=\linewidth]{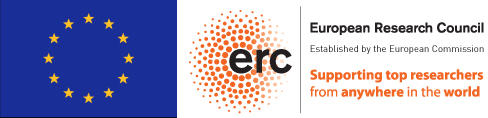}}}
\end{minipage}\hspace{1.2em}%
\begin{minipage}[t]{\dimexpr\columnwidth-3.2cm-0.8em\relax}
  This work was conducted as part of the AI-Twin project, which is funded by the
  European Research Council (ERC-2024-ADG) as part of the European Union's
  Horizon 2020 research and innovation program (grant agreement no.\ 101200584).
\end{minipage}
\end{acks}

\bibliographystyle{ACM-Reference-Format}
\bibliography{ahs_bibliography}

\appendix

\end{document}